\documentclass[submitting]{nst}

\usepackage{subfigure,dcolumn}
\usepackage{epstopdf}
\usepackage{mhchem}
\usepackage{upgreek}
\usepackage{longtable}
\usepackage{makecell,rotating,multirow,diagbox}
\usepackage{changes}
\usepackage{longtable}
\usepackage{supertabular}
\usepackage{CJK, upgreek, fancyhdr}
\usepackage{lastpage}
\usepackage{longtable}
\usepackage{changes}
\usepackage{etoolbox}
\usepackage{setspace}
\usepackage{tabularx}
\usepackage{hyperref}
\usepackage{supertabular}
\usepackage[normalem]{ulem}
\usepackage{eqnarray}
\usepackage{booktabs} 
\usepackage{subfigure}
\usepackage{amsmath}
\usepackage{graphicx}
\usepackage{float} 
\usepackage{subfigure}
\usepackage[thicklines]{cancel}
\usepackage{ulem}
\usepackage{setspace}
\usepackage{graphicx}

\begin{document}

\title{Charge state regulation of nuclear excitation by electron capture in $^{229}$Th ions}
\thanks{This work was supported by the National Natural Science Foundation of China (Grant No.12135009, 12375244), the Natural Science Foundation of Hunan Province of China (Grant No.2025JJ30002), and the Innovation Foundation for Postgraduate (Grant No.XJQY2024046, XJJC2024063).}

\author{Yang-Yang Xu}
\affiliation{College of Science, National University of Defense Technology, 410073 Changsha, People's Republic of China}

\author{Qiong Xiao}
\affiliation{College of Science, National University of Defense Technology, 410073 Changsha, People's Republic of China}

\author{Jun-Hao Cheng}
\affiliation{College of Science, National University of Defense Technology, 410073 Changsha, People's Republic of China}

\author{Wen-Yu Zhang}
\affiliation{College of Science, National University of Defense Technology, 410073 Changsha, People's Republic of China}

\author{Tong-Pu Yu}
\email[Corresponding author, ]{tongpu@nudt.edu.cn}
\affiliation{College of Science, National University of Defense Technology, 410073 Changsha, People's Republic of China}

\begin{abstract}
Nuclear excitation by electron capture (NEEC) in $^{229}$Th holds significant potential for precise nuclear state manipulation. In this study, we thoroughly  investigate NEEC in $^{229}\text{Th}^{q+}$ ions by integrating quantum numbers ($n, l, j$) effects and analyzing key parameters (e.g., resonance energy $E_r$, cross section $\sigma$, resonance strength $S$, and NEEC transition width $\Gamma_{\text{NEEC}}$) influences across charge state from $q=1^+$ to $90^+$. Especially, we focus on the charge-state regulation of the isomeric state (IS, 8.36 eV) and second-excited state (SE, 29.19 keV). Our calculations uncover critical charge-state-dependent behaviors of NEEC in $^{229}\text{Th}$ ions: (1) For the IS, valid NEEC channels exhibit threshold migration, where the dominant principal quantum number $n$ increases linearly with $q$ following the relation $n \approx 1.28q + 4.23$; meanwhile, single-$n$-channel $S$ stabilizes between $10^{-2}$ to $10^0$ barn eV via compensatory nucleus-electron coupling, ensuring the total resonance $S$ constant. (2) For the SE, its excitation energy far exceeds nearly all electron binding energies, leading to negligible channel screening and  causing the total $S$ to increase monotonically with $q$. This research clarifies the intrinsic mechanisms of charge-state-driven nuclear-electronic interactions in $^{229}\text{Th}^{q+}$ NEEC and provides a critical reference for future experimental efforts to manipulate $^{229}\text{Th}$ nuclear states, particularly via indirect regulation of the SE.
\end{abstract}


\maketitle

\section{Introduction}
\label{Sec.I}

The nucleus of $^{229}$Th has long stood out in low-energy nuclear physics and precision spectroscopy due to its unique low-lying isomeric state (IS), with an excitation energy of approximately 8.35574(3) eV above the ground state \cite{PhysRevLett.132.182501}. This is the lowest known nuclear excited state to date, bridging the traditionally distinct realms of nuclear physics (e.g., nuclear structure, excitation energies) and atomic physics (e.g., electron binding, charge-state effects) \cite{Thirolf2024,PhysRevA.109.042806,PhysRevA.95.032503,PhysRevLett.127.052501,PhysRevC.106.024606}. Such a peculiar property makes $^{229}$Th a prime candidate for next-generation nuclear optical clocks \cite{PhysRevLett.108.120802,Thirolf_2019}, a sensitive probe for testing fundamental constant variations \cite{PhysRevLett.104.200802,PhysRevLett.97.092502,PhysRevLett.102.210801,PhysRevA.102.052833}, and a platform for exploring nuclear-laser interactions\cite{PhysRevC.110.054307,NuclSciTech.20.36,PhysRevA.99.013422,PhysRevLett.106.162501,PhysRevLett.130.112501,PhysRevC.110.L051601,10.1063/1.5093535}.

To harness these applications, efficient and controllable population of the $^{229}$Th isomeric state is paramount. Early approaches relied on natural decay processes, such as the $\alpha$ decay of ${}^{233}\mathrm{U}$ \cite{PhysRevLett.98.142501,PhysRevC.94.014302} or the $\beta$ decay of ${}^{229}\mathrm{Ac}$ \cite{PhysRevC.100.024315}. However, these methods suffer from low yields, fixed branching ratios, and uncontrollable recoil energies, limiting their practicality. Alternative strategies, including direct laser excitation and indirect pumping via the 29189.93(0.07) eV second-excited state (SE), have been attempted, but they face challenges such as inefficient coupling between photons and nuclear states or narrow resonance conditions that are difficult to satisfy experimentally \cite{masuda2019x,10.1063/5.0251667}.

Against this backdrop, nuclear excitation by electron capture (NEEC) has emerged as a promising mechanism for populating nuclear isomers. Proposed almost half a century ago \cite{GOLDANSKII1976393}, NEEC describes a resonant process where a free electron is captured into an atomic orbital, with the released energy selectively exciting the nucleus to a higher energy state. It is effectively the time-reversed counterpart of internal conversion (IC) \cite{PhysRevLett.118.042501,PhysRevC.101.054602}. Theoretically, NEEC offers advantages such as tunable resonance energies and potentially high excitation cross sections \cite{CUE198925,PhysRevC.47.323,PhysRevA.73.012715,PhysRevLett.99.172502,PhysRevLett.120.052504,PhysRevLett.128.212502,10.3389/fphy.2023.1203401,10.3389/fphy.2024.1410076,PhysRevC.108.L031302,10.1063/5.0212163,PhysRevC.110.014330,NuclSciTech2210-3147}, making it a viable candidate for controlled isomer excitation. Despite its potential, experimental verification of NEEC remains elusive and controversial. The 2018 claim of observing NEEC in $^{93\text{m}}$Mo sparked intense debate \cite{chiara2018isomer}, as subsequent theoretical calculations showed a discrepancy of up to nine orders of magnitude with the reported signal strength \cite{PhysRevLett.122.212501}. Follow-up experiments using isomer beams failed to reproduce the effect \cite{PhysRevLett.128.242502}, highlighting the need for more rigorous theoretical frameworks and optimized experimental designs to clarify the NEEC process.

In the specific context of $^{229}$Th, NEEC's role in exciting the isomeric state remains largely unexplored. Previous studies have focused on electronic bridge (EB) \cite{PhysRevA.81.042516,PhysRevLett.105.182501,Bilous_2018,PhysRevC.100.044306,PhysRevC.102.024604,PhysRevC.106.064608,PhysRevLett.124.192502,NuclSciTech10.1107}, electron transition (NEET) \cite{Dzyublik2011,PhysRevC.88.054616,TKALYA1992209,SAKABE20051}, inelastic electron scattering (NEIES) \cite{PhysRevLett.124.242501,PhysRevC.106.044604,PhysRevC.106.064604,PhysRevC.110.064621}, or laser-driven recollision processes \cite{PhysRevLett.127.052501,PhysRevC.106.024606}, but the potential of NEEC, particularly its dependence on charge states, electron configurations and resonance conditions, has not been systematically addressed. Key questions persist: How does varying charge states of $^{229}$Th ions modulate the NEEC efficiency? What are the optimal experimental parameters to observe NEEC-induced isomer excitation in $^{229}$Th? In this work, we aim to address these unresolved aspects by providing a comprehensive analysis of NEEC in $^{229}$Th ions. By investigating the interplay between charge states, electron capture channels, and nuclear transition energies, we quantify NEEC cross sections and resonance strengths under various conditions. Our results clarify certain microscopic mechanisms of nuclear-electron coupling in $^{229}$Th and provide specific references for experimental attempts to verify NEEC, which may contribute to understanding this process and its potential in manipulating the $^{229}$Th isomeric state.

This article is structured as follows. The next section outlines the theoretical framework for NEEC calculations. In Section \ref{Sec.III}, the detailed results and discussions on the roles of quantum numbers and charge states are provided. In Section \ref{Sec.IV}, we provide a brief summary.

\section{Theoretical framework}\label{Sec.II}

To quantitatively describe the excitation of the $^{229}$Th nucleus to its excited state (IS and SE) via NEEC, we establish a theoretical framework centered on the interaction between free electrons and the nucleus, where the electron is captured into a bound state while transferring energy to excite the nucleus.

The core of NEEC lies in the coupling between the electron and the nucleus, mediated by both electromagnetic currents and Coulomb interactions. The interaction Hamiltonian $H_{\mathrm{int}}$ is expressed as \cite{RevModPhys.30.353}
\begin{equation}
	\label{eq1}
	\begin{aligned}
		H_{\mathrm{int}}= & -\frac{1}{c} \int\left[\boldsymbol{j}_{\mathrm{n}}(\boldsymbol{r})+\boldsymbol{j}_{\mathrm{e}}(\boldsymbol{r})\right] \cdot \mathbf{A}(\boldsymbol{r}) d \tau \\
		& +\int \frac{\rho_{\mathrm{n}}(\boldsymbol{r}) \rho_{\mathrm{e}}\left(\boldsymbol{r}^{\prime}\right)}{\left|\boldsymbol{r}-\boldsymbol{r}^{\prime}\right|} d \tau d \tau^{\prime}.
	\end{aligned}
\end{equation}
Here, the first term describes the coupling between nuclear current density $\boldsymbol{j}_{\mathrm{n}}(\boldsymbol{r})$ and electron current density $\boldsymbol{j}_{\mathrm{e}}(\boldsymbol{r})$ via the radiation field’s vector potential $\mathbf{A}(\boldsymbol{r})$. The second term accounts for the Coulomb interaction between nuclear charge density $\rho_{\mathrm{n}}(\boldsymbol{r})$ and electron charge density $\rho_{\mathrm{e}}\left(\boldsymbol{r}^{\prime}\right)$.

For the electronic states involved in NEEC, the initial state is a free electron with kinetic energy $E_i$, and the final state is a bound atomic orbital with energy $E_f$. Energy conservation requires $E_i = \Delta E + |E_f|$, where $\Delta E$ is the energy difference between the final and initial nuclear states. Electron wavefunctions are solutions to the Dirac equation under the Dirac-Hartree-Fock-Slater (DHFS) potential \cite{PhysRev.92.978}
\begin{equation}
	\label{eq2}
	\begin{aligned}
		V_{\mathrm{DHFS}}(r)=V_{\mathrm{n}}(r)+V_{\mathrm{e}}(r)+V_{\mathrm{ex}}(r),
	\end{aligned}
\end{equation}
where the nuclear potential $V_{\text {n}}(r)$ arises from a Fermi-distributed nuclear charge density \cite{PhysRev.101.1131}, the electronic potential $V_{\mathrm{e}}(r)$ from the electron cloud, and the exchange potential $V_{\mathrm{ex}}(r)$ from a piecewise Thomas-Fermi approximation for exchange interactions \cite{PhysRev.81.385}.

The NEEC cross section, which quantifies the probability of the process, is given by
\begin{equation}
	\label{eq3}
	\begin{aligned}
		\sigma_{\mathrm{E(M) \lambda}} = &\frac{4\pi^2}{c^2}\frac{E_i + c^2}{p_i^3} \times \sum_{l_i, j_i, l_f, j_f} \frac{\kappa^{2 \lambda+2}}{(2 \lambda-1)!!^2} \\
		& \times \frac{\left(2 l_i+1\right)\left(2 l_f+1\right)\left(2 j_i+1\right)\left(2 j_f+1\right)}{(2 \lambda+1)^2} \\
		& \times\left(\begin{array}{ccc}
			l_f & l_i & \lambda \\
			0 & 0 & 0
		\end{array}\right)^2\left\{\begin{array}{ccc}
			l_i & \lambda & l_f \\
			j_f & 1 / 2 & j_i
		\end{array}\right\}^2\left|T_{f i}^{\mathrm{E(M)\lambda}}\right|^2
		\\&\times B\left(\mathrm{E(M) \lambda}, I_i \rightarrow I_f\right) 
		\frac{\Gamma_{\mathrm{NEEC}}}{\Delta E^2+\Gamma_{\mathrm{NEEC}}^2 / 4}.
	\end{aligned}
\end{equation}
In this expression, E and M denote electric and magnetic multipole transitions, respectively, with $\lambda$ indicating the order of the multipole, while $B(\text{E(M)}\lambda, I_i \to I_f)$ represents the reduced nuclear transition probability that encapsulates the intrinsic strength of the nuclear transition. $\kappa=E_{exc}/c$ with $E_{exc}$ being the excitation energy and $T_{fi}^{\text{E(M)}\lambda}$ are the radial matrix elements that quantify the overlap between the initial and final electronic wavefunctions, a critical factor in determining transition probabilities. $\Gamma_{\text{NEEC}}$ represents the total transition width, which is the sum of $\Gamma_{\text{spon}}$ and $\Gamma_{\text{n}}$. $\Gamma_{\text{spon}}$ corresponds to the spontaneous emission width of the final electronic state, reflecting the probability of the electron transitioning to lower energy levels through photon emission. $\Gamma_{\text{n}}$ denotes the nuclear width, which includes $\Gamma_{\text{IC}}$ (internal conversion width, characterizing the probability of energy transfer from the nucleus to atomic electrons) and $\Gamma_\gamma$ (gamma decay width, describing the probability of radiative decay via $\gamma$-ray emission).

The radial matrix elements \(T_{fi}^{\text{E}\lambda}\) and \(T_{fi}^{\text{M}\lambda}\) are defined as
\begin{equation}
	\label{eq4}
	\begin{aligned}
		&T_{f i}^{\text{E} \lambda}= \int_0^{\infty} h_\lambda^{(1)}(\kappa r)\left[g_i(r) g_f(r)+f_i(r) f_f(r)\right] r^2 dr\\
		& -\int_0^{\infty} h_{\lambda-1}^{(1)}(\kappa r)\frac{\kappa}{\lambda}\left[g_i(r) g_f(r)+f_i(r) f_f(r)\right] r^3 d r,
	\end{aligned}
\end{equation}
and
\begin{equation}
	\label{eq5}
	\begin{aligned}
		&T_{f i}^{\text{M} \lambda}=\frac{\eta_i+\eta_f}{\lambda} \\&\times \int_0^{\infty} h_\lambda^{(1)}(\kappa r) \left[g_i(r) f_f(r)+g_f(r) f_i(r)\right] r^2 d r.
	\end{aligned}
\end{equation}
Here, $h^{(1)}(\kappa r)$ is the spherical Hankel function of the first kind, and $\eta=(l-j)(2j+1)$ with $l$ replaced by $l^{\prime}=2j-l$ for $\eta<0$ or M$\lambda$ transitions. $g(r)$ and $f(r)$ denote the radial components of the upper and lower parts of the Dirac spinor \cite{SALVAT2019165}, respectively, with the subscripts $i$ and $f$ corresponding to the initial and final states of the electron.

When the incoming electron energy exhibits a distribution, the resonance strength $S$ serves to simplify calculations
\begin{equation}
	\label{eq6}
	\begin{aligned}
		S=\int \sigma_{\text{NEEC}}\left( E _i\right)d E_i .
	\end{aligned}
\end{equation}

All calculations above use atomic units (a.u.) with the reduced Planck constant ($\hbar$) = electron mass ($m_e$) = elementary charge ($e$) = 1. Lengths measured in Bohr radii, energies in Hartrees, and the speed of light given by $c = 1/\alpha \approx 137.036$ where $\alpha$ is the fine-structure constant.

\section{Results and discussion}\label{Sec.III}

In this study, we initiate our exploration of the NEEC process in singlely charged thorium ions ($\text{Th}^{1+}$) with the ground-state electronic configuration $[\text{Rn}] 6d^{1}7s^{2}$. For the $^{229}\text{Th}$ nucleus, three primary states exist with distinct spin-parity configurations including the ground state (GS) of 5/2$^{+}$, IS of 3/2$^{+}$, and SE of 5/2$^{+}$. Transitions between the GS and both excited states adhere to the selection rules governing multipole radiation and parity conservation, allowing them to proceed via magnetic dipole (M1) or electric quadrupole (E2) modes. Theoretical calculations have established the reduced transition probabilities with the IS-to-GS transition exhibiting a $B(\text{E}2)$ value of 27 W.u. and a $B(\text{M}1)$ value of 0.0076 W.u. while the SE-to-GS transition shows a $B(\text{E}2)$ value of 39.49 W.u. and a $B(\text{M}1)$ value of 0.0043 W.u. \cite{PhysRevLett.118.212501,PhysRevLett.122.162502,PhysRevC.103.014313,PhysRevC.105.064313}. With this background, we aim to clarify how the principal quantum number ($n$) of the final-state orbit affects the $\sigma$, $\Gamma_{\text{NEEC}}$ and $S$, and to quantify the contributions of M1 and E2 transitions. To this end, we focus on the $s_{1/2} \to nd_{3/2}$ channels, as they support both transition types.

\begin{figure}[H]
	\includegraphics[width=10.25cm,height=7.3cm]{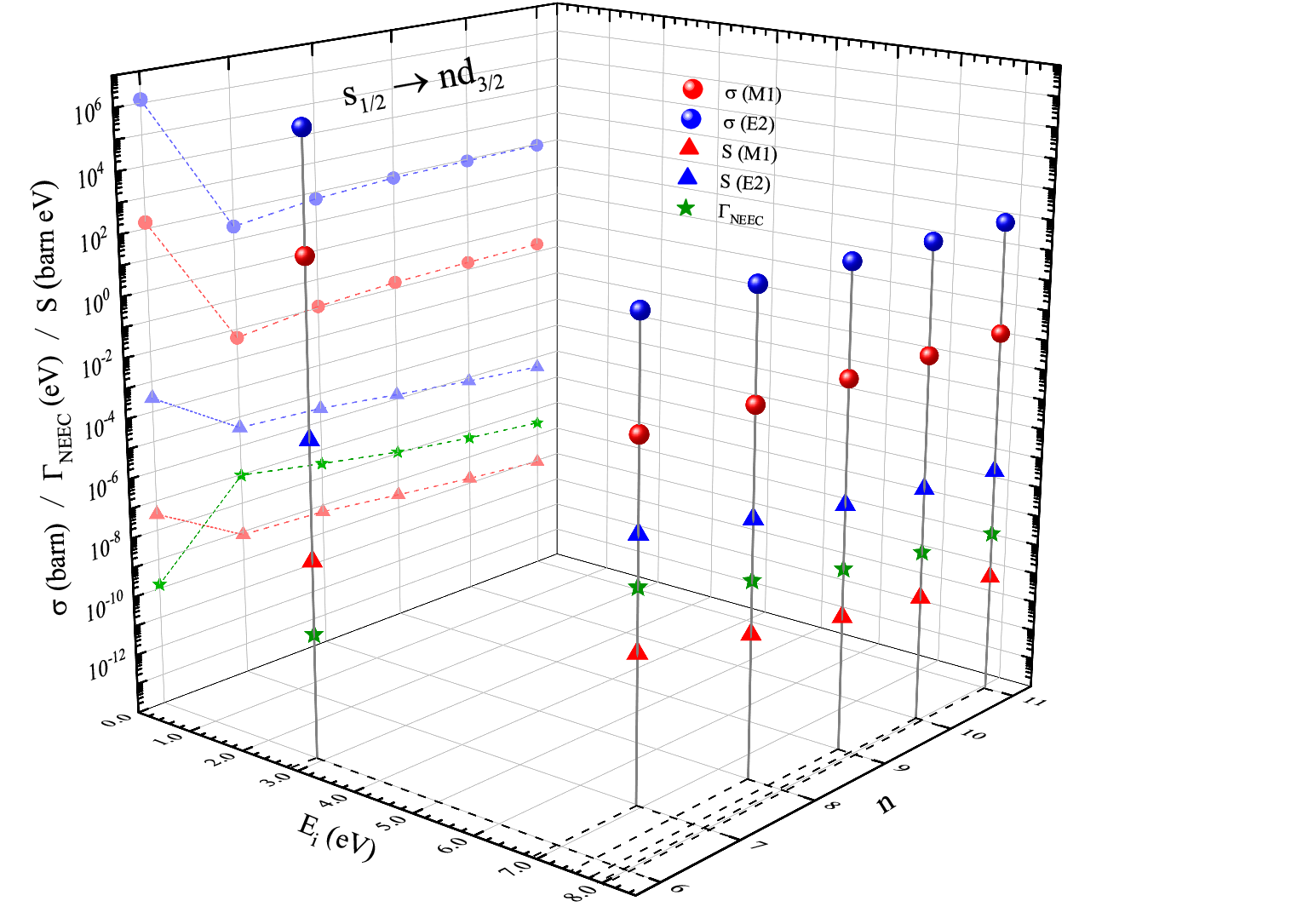}
	\caption{Three-dimensional visualization of NEEC parameters $\sigma$, $\Gamma_{\text{NEEC}}$, $S$ ($z$-axis) as functions of $E_i$ ($x$-axis) and final-state $n$ ($y$-axis), with projections on the $y-z$ plane for clarity.}
	\label{fig 111}
\end{figure}

\begin{figure}[H]
	\includegraphics[width=10.25cm,height=7.3cm]{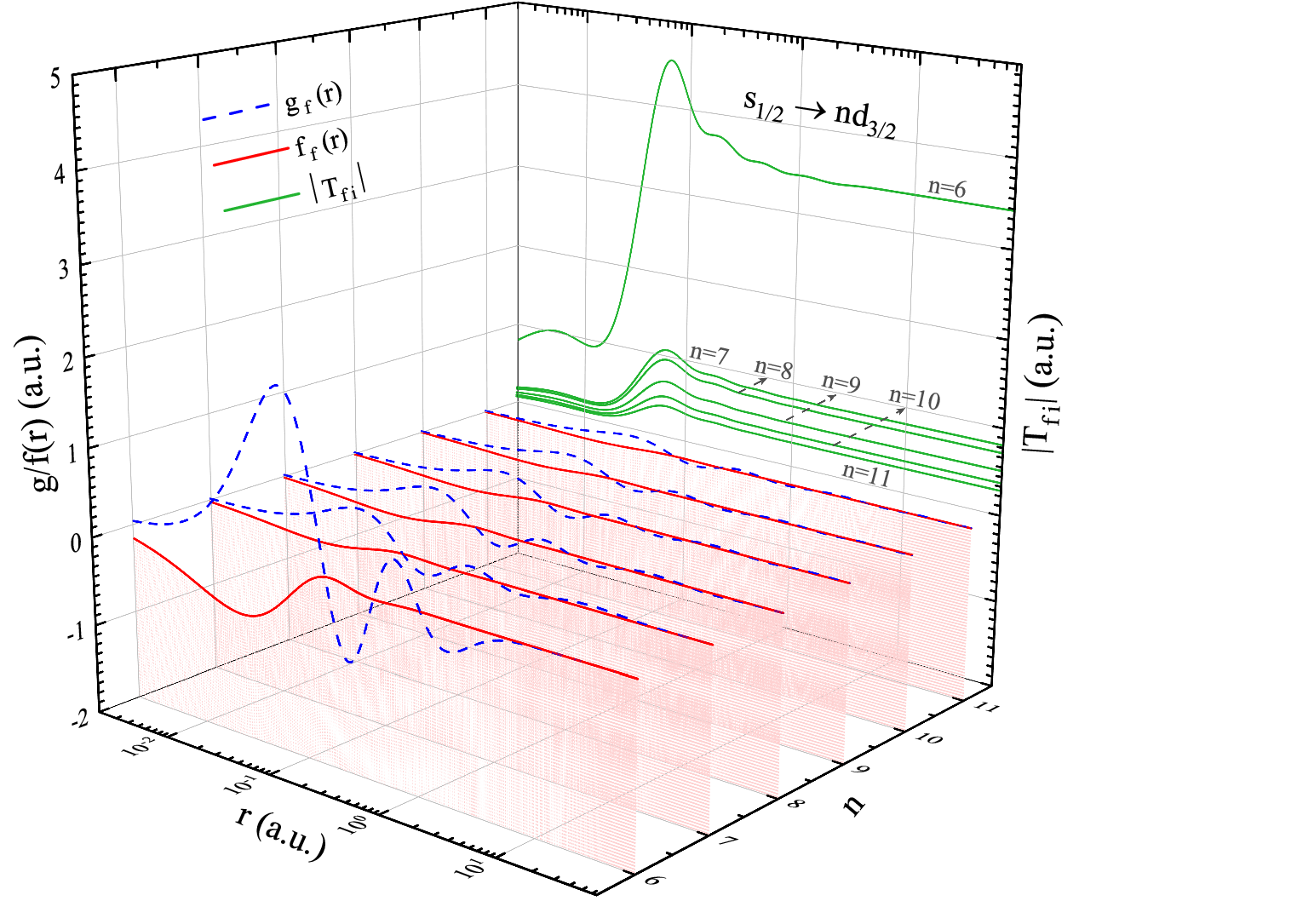}
	\caption{Radial components of Dirac spinors ($g(r)$, $f(r)$) and transition matrix elements for different final states.}
	\label{fig 222}
\end{figure}

The calculation results are shown in Fig. \ref{fig 111}. For systems with spontaneous radiation channels, $\Gamma_{\text{NEEC}}$ is dominated by $\Gamma_{\text{spon}}$ as $\Gamma_{\text{IC}}$ is typically orders of magnitude smaller. This is evident through the abrupt rise in $\Gamma_{\text{NEEC}}$ between the $6d_{3/2}$ and $7d_{3/2}$ states, as observed from projections on the $y$-$z$ plane. The $6d_{3/2}$ state has limited spontaneous radiation channels, with only the $\text{E}1$ transition to $5f_{5/2}$, whereas the $7d_{3/2}$ state possesses five such pathways. As $n$ increases from 7 to 11, despite more radiative channels becoming available, $\Gamma_{\text{NEEC}}$ decreases slowly. This trend arises because larger $n$ pushes the electron orbit further from the nucleus, weakening the electron-nucleus interactions and reducing the probability of spontaneous radiation.

From the cross-section formula Eq. (\ref{eq3}), the magnitude of the cross-section peak is jointly determined by $E_i$, $\Gamma_{\text{NEEC}}$ and $|T_{fi}|$, which is inversely proportional to the first two and directly proportional to the square of the latter. The resonance center energy $E_r$ rises with $n$ as higher-$n$ orbitals own weaker electron binding energies and thus demand greater initial electron kinetic energy to maintain energy conservation. To identify the primary determinant of the cross-section peak, we also plot the radial components $g(r)$ and $f(r)$ and the corresponding $|T_{fi}|$ for each final state in Fig. \ref{fig 222}. With increasing $n$, the lower amplitude in the excitation-effective region grows, leading to a smaller $|T_{fi}|$ that decays distinctly: dropping sharply from $n$=6 to $n$=7 and then decreasing much more slowly for $n$$\geq$7. This pattern aligns with the $\sigma$ trend in Fig. \ref{fig 111}, where the initial decline arises because the large reduction in $|T_{fi}|$ and the increase in $\Gamma_{\text{NEEC}}$ outweigh the rising $E_i$. For $n$>7, the rising $E_i$ and slower decay of both $|T_{fi}|$ and $\Gamma_{\text{NEEC}}$ collectively drive $\sigma$ to increase. Furthermore, as $n$ increases, $S$ shows a stable downward trend. Despite fluctuations in $\sigma$, the combined effects of diminishing radial wavefunction overlap and the gradual reduction in $\Gamma_{\text{NEEC}}$ dominate, resulting in an overall decrease in $S$. In terms of transition contributions, E2 transitions play a dominant role, with their $\sigma$ and $S$ far exceeding those of M1 transitions. This disparity arises from transition selection rules and orbital angular momentum characteristics that favor E2 transitions in this channel, making them the primary contributors to the NEEC process.

\begin{table}[htbp]
	\caption{The calculation results for NEEC correspond to the E2 transitions of the ground-state $\text{Th}^{1+}$ with a final-state $n = 8$.}
	\label{tab111}
	\centering
	\setlength{\tabcolsep}{4.8pt}
	\begin{tabular}{|l|c|c|c|}
		\hline
		\multicolumn{1}{|c|}{Transition} & \multicolumn{1}{c|}{\( \sigma \) (barn)} & \multicolumn{1}{c|}{\( S \) (barn eV)} & \multicolumn{1}{c|}{\( \Gamma_{\text{NEEC}} \) (eV)} \\
		\hline
		$s_{1/2} \to 8d_{3/2}$ & 5.84E+01 & 2.98E-06 & 3.19E-08 \\
		$s_{1/2} \to 8d_{5/2}$ & 1.41E-01 & 7.50E-09 & 3.33E-08 \\
		$p_{1/2} \to 8p_{3/2}$ & 7.70E+04 & 7.44E-04 & 6.04E-09 \\
		$p_{1/2} \to 8f_{5/2}$ & 4.69E-03 & 2.55E-10 & 3.40E-08 \\
		$p_{3/2} \to 8p_{1/2}$ & 1.25E+05 & 7.89E-04 & 3.94E-09 \\
		$p_{3/2} \to 8p_{3/2}$ & 4.30E+04 & 4.16E-04 & 6.04E-09 \\
		$p_{3/2} \to 8f_{5/2}$ & 4.75E-04 & 2.58E-11 & 3.40E-08 \\
		$p_{3/2} \to 8f_{7/2}$ & 9.09E-04 & 4.78E-11 & 3.28E-08 \\
		$d_{3/2} \to 8s_{1/2}$ & 8.79E+01 & 8.45E-06 & 6.01E-08 \\
		$d_{3/2} \to 8d_{3/2}$ & 1.65E+02 & 8.39E-06 & 3.19E-08 \\
		$d_{3/2} \to 8d_{5/2}$ & 4.19E+01 & 2.23E-06 & 3.33E-08 \\
		$d_{5/2} \to 8s_{1/2}$ & 8.52E-02 & 8.19E-09 & 6.01E-08 \\
		$d_{5/2} \to 8d_{3/2}$ & 4.47E+01 & 2.28E-06 & 3.19E-08 \\
		$d_{5/2} \to 8d_{5/2}$ & 1.59E+02 & 8.49E-06 & 3.33E-08 \\
		$f_{5/2} \to 8p_{1/2}$ & 1.23E-01 & 7.72E-10 & 3.94E-09 \\
		$f_{5/2} \to 8p_{3/2}$ & 2.91E-04 & 2.81E-12 & 6.04E-09 \\
		$f_{5/2} \to 8f_{5/2}$ & 1.15E+00 & 6.23E-08 & 3.40E-08 \\
		$f_{5/2} \to 8f_{7/2}$ & 1.85E-01 & 9.71E-09 & 3.28E-08 \\
		$f_{7/2} \to 8p_{3/2}$ & 2.78E-02 & 2.68E-10 & 6.04E-09 \\
		$f_{7/2} \to 8f_{5/2}$ & 1.64E-01 & 8.89E-09 & 3.40E-08 \\
		$f_{7/2} \to 8f_{7/2}$ & 1.42E+00 & 7.46E-08 & 3.28E-08 \\
		\hline
	\end{tabular}
\end{table}

\begin{figure}[H]
	\includegraphics[width=10.25cm,height=7.5cm]{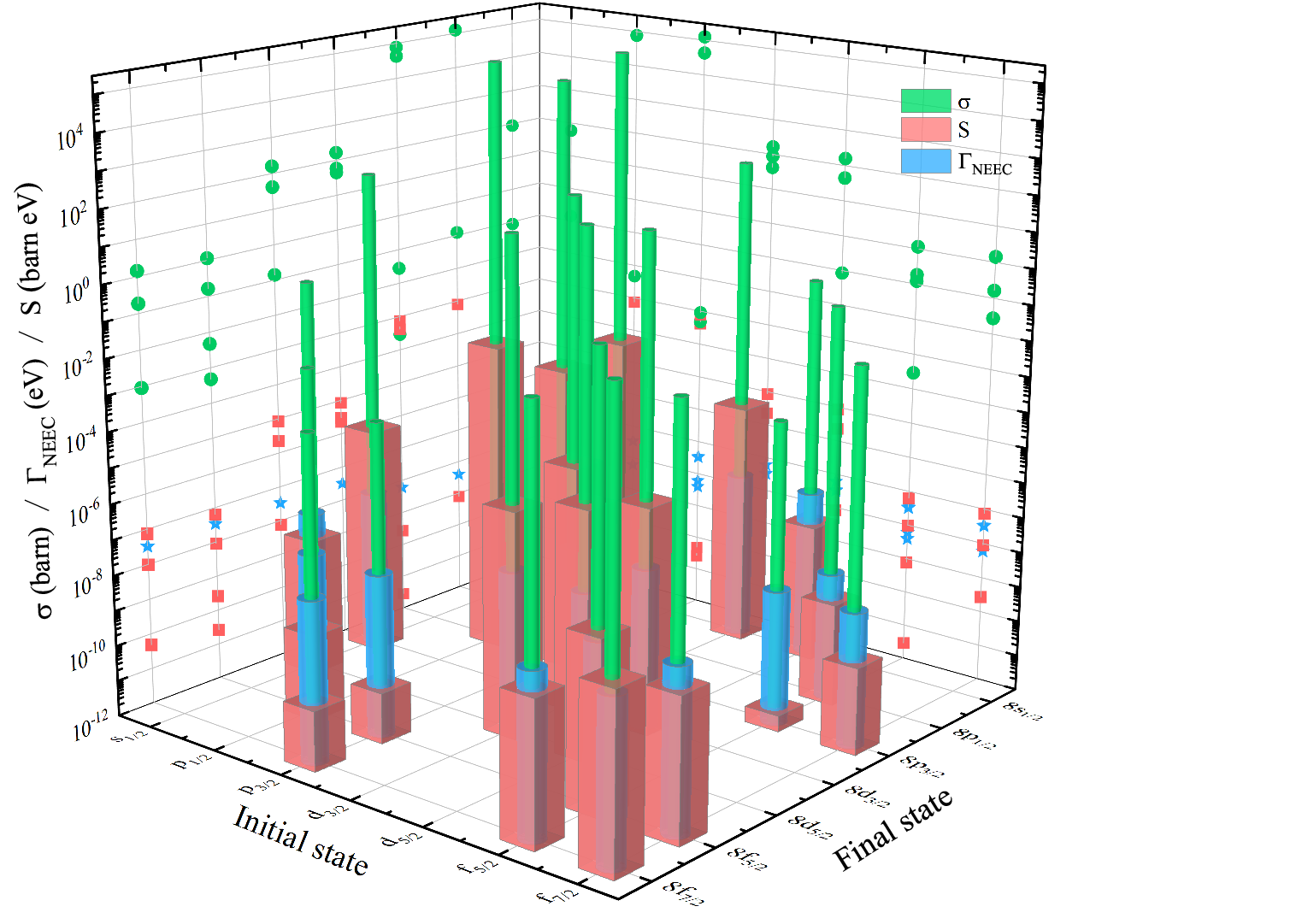}
	\caption{Three-dimensional visualization of NEEC parameters for ground-state $\text{Th}^{1+}$: influence of $l$ and $j$ in $n$=8 transitions (with side-plane projections; $x$/$y$-axes: initial/final states; $z$-axis: $\sigma$, $S$, and $\Gamma_{\text{NEEC}}$).}
	\label{fig 333}
\end{figure}

To further investigate the influence of the orbital angular momentum $l$ and total angular momentum $j$ on the NEEC parameters, we focused on E2 transitions in ground-state $\text{Th}^{1+}$ with final-state $n$=8. The results are presented in Table \ref{tab111} and visualized in Fig. \ref{fig 333}, in which side-plane projections show that $\sigma$ and $S$ decrease overall with increasing $l$. This arises from reduced radial wavefunction overlap between initial and final states with higher $l$. Electron clouds diverge more significantly, suppressing transition probability and attenuating $\sigma$. At the microscopic level, transitions with matched $l$ (e.g., $p \to p$, $f \to f$) exhibit local maxima in $\sigma$ due to improved symmetry matching, whereas mismatched $l$ further diminishes the radial wavefunction overlap and $\sigma$. Resonance strength $S$, jointly determined by $\sigma$ and $\Gamma_{\text{NEEC}}$, closely tracks the trend of $\sigma$ owing to the modest magnitude of $\Gamma_{\text{NEEC}}$'s increase with $l$. For fixed $l$, $\Gamma_{\text{NEEC}}$ exhibits weak sensitivity to $j$, with values for transitions sharing the same $l$ but differing $j$ remaining within the same order of magnitude. In contrast, $j$ strongly modulates $\sigma$: for example, $d$-orbital transitions with different $j$ exhibit cross-section variations up to three orders of magnitude, underscoring $j$'s role in shaping wavefunction overlap with the initial state. 

\begin{figure}[H]
	\centering
	\includegraphics[width=9.3cm,height=6.0cm]{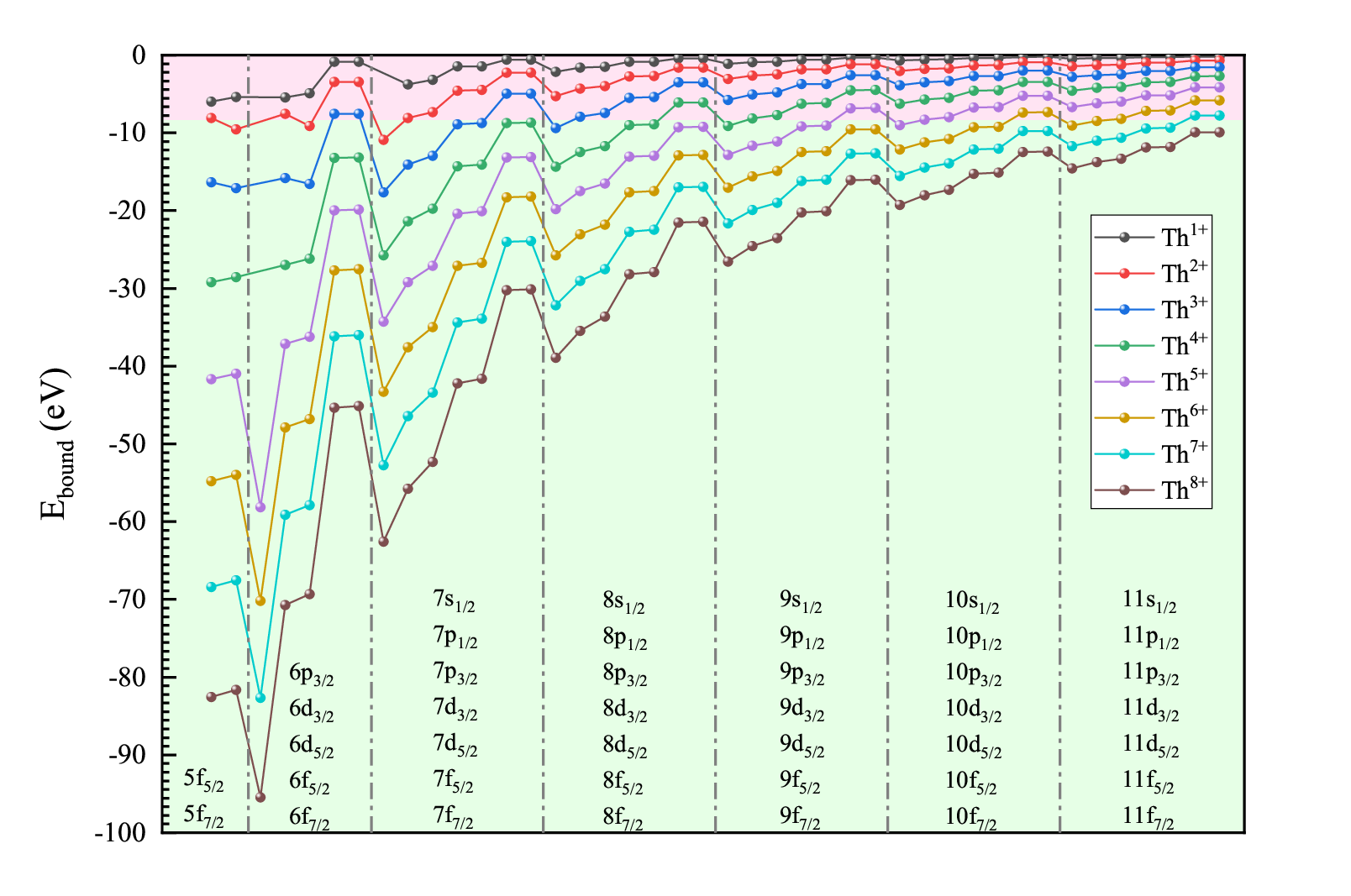}
	\caption{Binding energy spectrum of electronic holes in $\text{Th}^{q+}$ across low charge states ($q=1^{+}\to8^{+}$). The pink region indicates valid NEEC channels with $|E_{\text{bound}}| < 8.36\ \text{eV}$.}
	\label{fig 444}
\end{figure}

\begin{figure}[H]
	\centering
	\includegraphics[width=8.5cm,height=4.9cm]{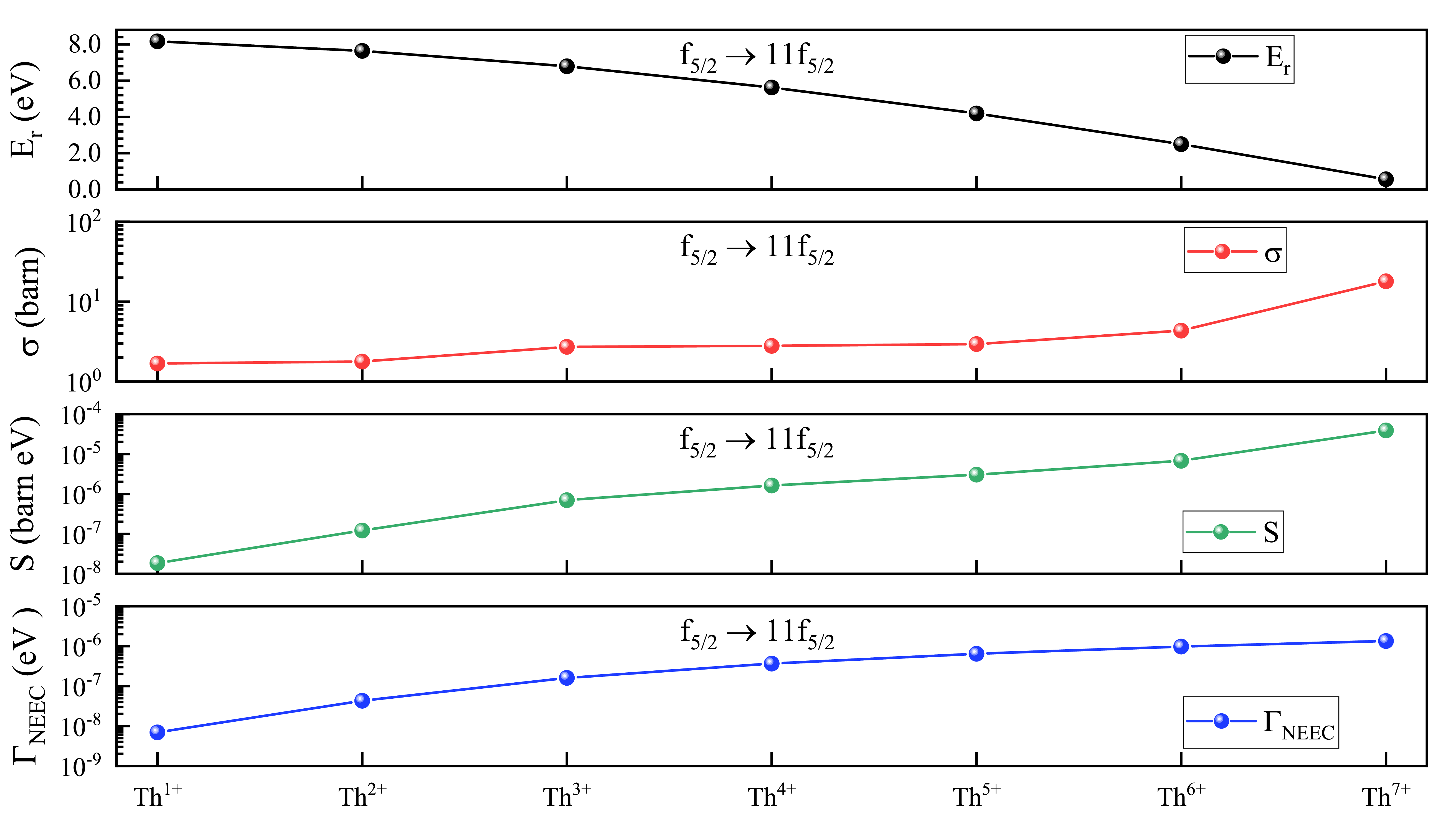}
	\caption{NEEC parameters ($E_r$, $\sigma$, $S$, $\Gamma_{\text{NEEC}}$) for the $f_{5/2} \to 11f_{5/2}$ transition as a function of charge state $q$.}
	\label{fig 555}
\end{figure}

The above results elucidate that quantum numbers constitute the core microscopic parameters governing the activity of NEEC channels, with the regulatory role of the ionic charge state $q$ manifesting precisely in its capacity to reshape the electronic binding energy spectrum and to screen these quantum numbers. Therefore, we proceed to explore the macroscopic regulation of $q$, building on the microscopic insights into quantum numbers. Fig. \ref{fig 444} illustrates the binding energy spectrum of electronic holes in low-$q$ and the pink region denotes valid IS NEEC channels with $|E_{\text{bound}}|<8.36\ \text{eV}$. As $q$ increases, the strengthened Coulomb potential elevates  $|E_{\text{bound}}|$, leaving fewer channels within the pink region—a process termed ``channel screening''. For instance, $\text{Th}^{1+}$ supports nearly all $5f$ to $11f$ holes, while $\text{Th}^{7+}$ retains only high-$n$, high-$l$ orbitals like $11f$. This reduces the number of viable channels, yet the remaining channels exhibit enhanced coupling strength, as revealed by Fig. \ref{fig 555}.

Focusing on the surviving $f_{5/2}$$\to$$11f_{5/2}$ channel, Fig. \ref{fig 555} details parameter evolution with $q$. While $E_r$ decreases monotonically, consistent with $E_r = 8.36 - |E_{\text{bound}}|$, $\sigma$ and $S$ exhibit significant increases. This apparent contradiction is resolved by considering the Coulomb contraction of electron wavefunctions in high-$q$ ions \cite{PhysRevC.110.064621}. Despite the large $n=11$ and $l=3$ values, which would typically disperse the wavefunction, the enhanced Coulomb potential pulls the wavefunction closer to the nucleus, amplifying $|T_{fi}|$ and thus increasing $\sigma$. Concurrently, $\Gamma_{\text{NEEC}}$ broadens due to stronger spontaneous radiation from tightly bound electrons, driving $S$ growth via the combined effect of rising $\sigma$ and $\Gamma_{\text{NEEC}}$.

\begin{figure}[H]
	\centering
	\includegraphics[width=8.4cm,height=6.5cm]{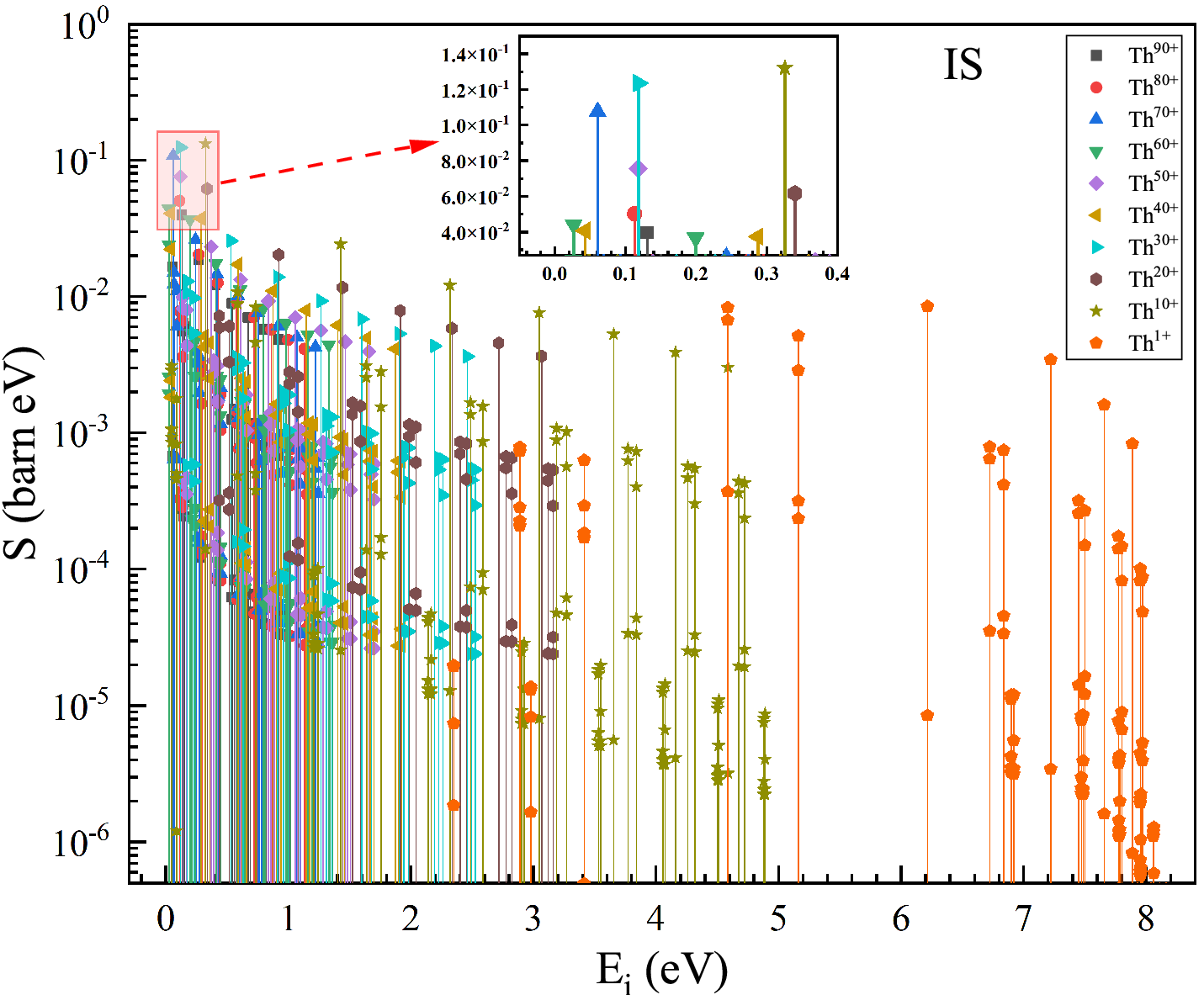}
	\caption{Multi-channel $S$ for the IS across charge states. Each color represents a distinct $\text{Th}^{q+}$ charge state.}
	\label{fig 666}
\end{figure}

To unravel the interplay between charge state and resonance dynamics, we first analyze the IS through its multi-channel behavior and underlying mechanisms. The multi-channel landscape of the IS, visualized in Fig. \ref{fig 666}, shows a key feature across all charge states ($q$=$1^+$, $10^+$, $\dots$, $90^+$): the strongest $S$ peaks cluster within an $E_r$ range of 0-0.4 eV, with intensities remaining remarkably consistent. This stability in energy distribution and strength arises from a balance between structural evolution and coupling enhancement. 

\begin{figure}[H]
	\centering
	\includegraphics[width=8.5cm,height=6.5cm]{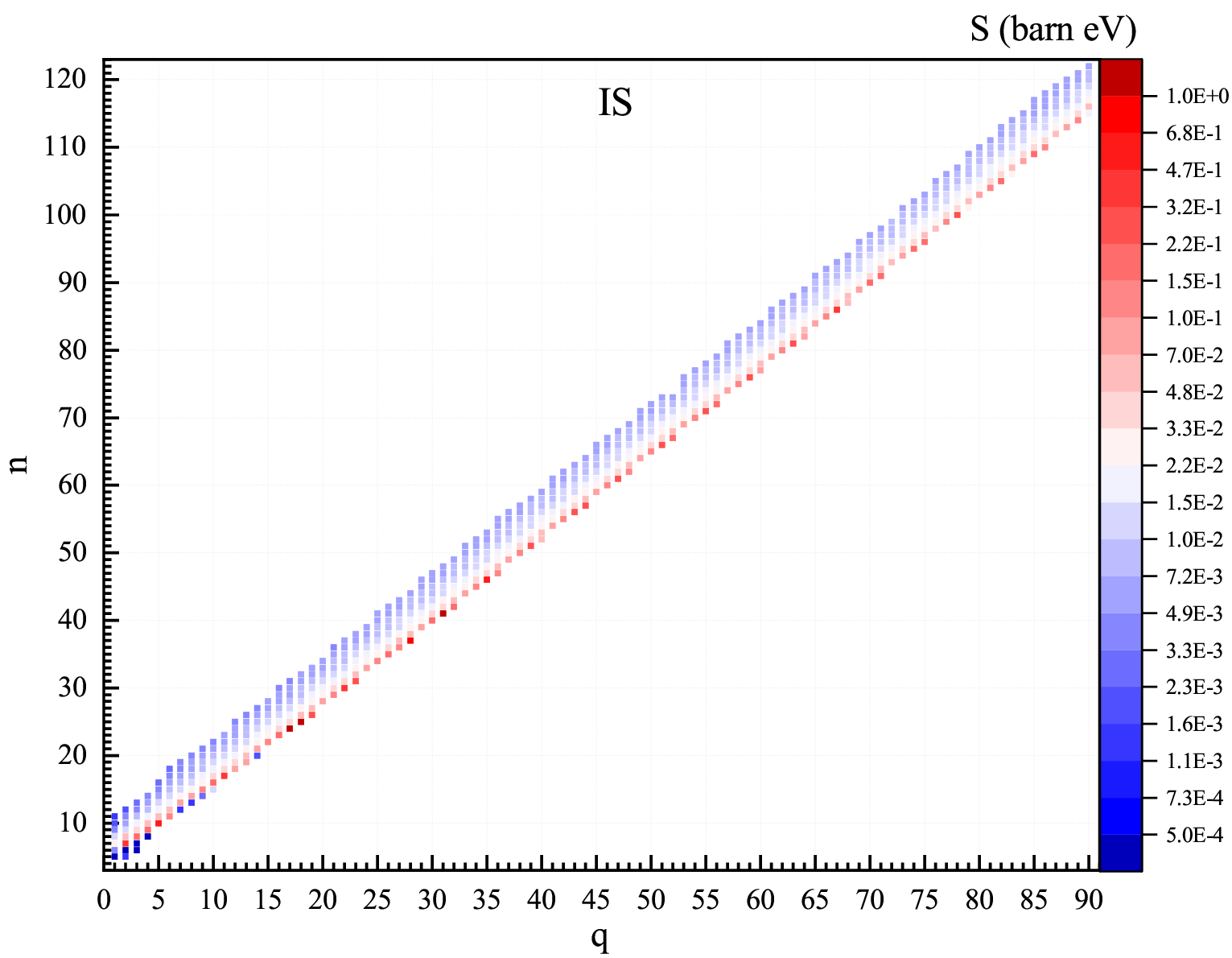}
	\caption{$S$ of IS NEEC channels as a function of $q$ and $n$.}
	\label{fig 777}
\end{figure}

Fig. \ref{fig 777} maps $S$ across $q$ (1$^+$$\to$90$^+$) and $n$ for the IS, which illustrates a striking threshold migration of valid NEEC channels with increasing $q$. At low $q$ (e.g., $q=1^{+}$), effective channels span a broad range of $n$>5, consistent with the abundant low-$q$ channels observed in Fig. \ref{fig 444}. As $q$ rises, this range shifts monotonically toward higher $n$ and for $q=90^{+}$, valid channels are restricted to $n$>115. This rigid migration directly arises from the need to maintain the energy-matching condition under a strengthened Coulomb potential—only high-$n$ orbitals, with their weaker binding energies, can satisfy this constraint. Quantitative analysis shows that the dominant $n$ (the $n$ contributing the maximum $S$ for each $q$) follows a linear relationship $n\approx$1.28$q$+4.23, confirming Coulomb potential as the key driver of orbital availability for IS. Notably, despite the significant shift in dominant $n$, the corresponding maximum $S$ exhibits remarkable stability across $q$, with values clustering between $10^{-5}$ $\text{barn eV}$ and $10^0$ $\text{barn eV}$. For example, $q=9^{+}$ (dominant $n$=15), $q=50^{+}$ (dominant $n$=65) and $q=89^{+}$ (dominant $n$=114) yield $S \approx 10^{-1}$ $\text{barn eV}$. This stability results from a compensatory interplay where Coulomb contraction amplifies $|T_{fi}|$ in high-$n$ orbitals, thereby offsetting the spatial dispersion inherent to larger $n$ values and balancing the reduced number of available channels with enhanced single-channel coupling strength.

\begin{figure}[H]
	\centering
	\includegraphics[width=8.4cm,height=6.5cm]{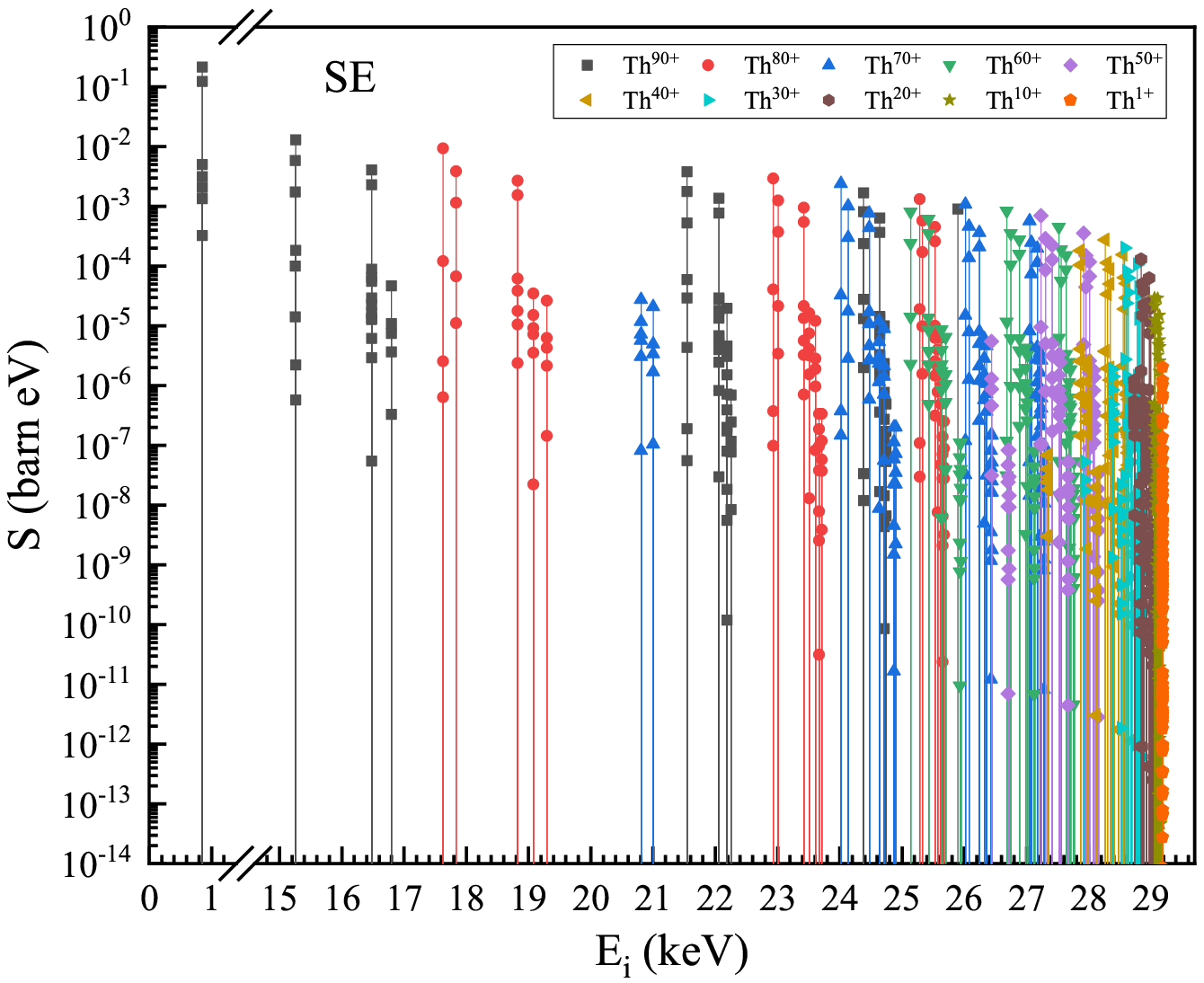}
	\caption{Multi-channel $S$ for the SE across charge states. Each color represents a distinct $\text{Th}^{q+}$ charge state.}
	\label{fig 888}
\end{figure}

Focus shifts to the SE, whose investigation is driven by its substantial role in indirectly populating the IS with approximately 90\% of SE decays occurring via transitions to the IS such that a high excitation rate of the SE would result in non-negligible indirect excitation of the IS \cite{masuda2019x}. The SE exhibits distinct behavior rooted in its unique energetic properties. Fig. \ref{fig 888} structured to mirror Fig. \ref{fig 666} for direct comparison shows two defining trends: maximum $S$ span a dispersed $E_r$ range, and their intensities surge with $q$ while $E_r$ itself decreases monotonically. These features reflect a liberation from the constraints governing the IS. 

\begin{figure}[H]
	\centering
	\includegraphics[width=8.5cm,height=6cm]{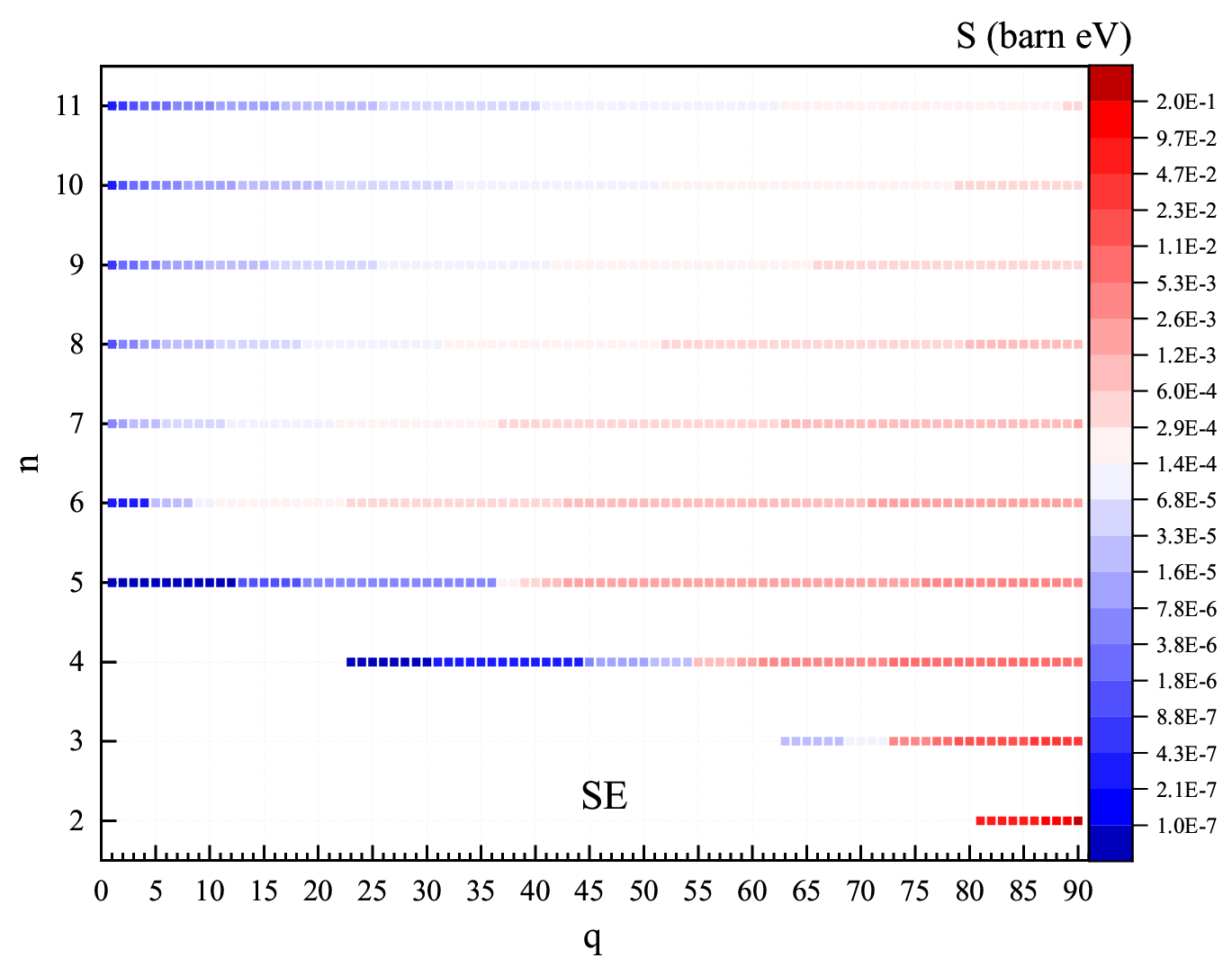}
	\caption{$S$ of SE NEEC channels as a function of $q$ and $n$.}
	\label{fig 999}
\end{figure} 

Fig. \ref{fig 999} shows the SE exhibits negligible threshold migration with valid channels confined to a narrow $n$ range ($n$>5 for low $q$, extending to $n$>2 at $q=90^{+}$) across all charge states. This insensitivity occurs because the SE excitation energy (29.19 keV) vastly exceeds electron binding energies for almost all $q$, obviating channel screening. Consequently, the maximum $S$ from the dominant $n$ for each $q$ grows monotonically by 4 orders of magnitude (e.g., from 4.65$\times 10^{-6}$ $\text{barn eV}$ at $q=1^{+}$ to 3.47$\times 10^{-1}$ $\text{barn eV}$ at $q=90^{+}$), driven solely by Coulomb contraction strengthening electron-nucleus coupling across all $n$ without countervailing channel loss.

\begin{figure}[H]
	\centering
	\includegraphics[width=8.7cm,height=6.1cm]{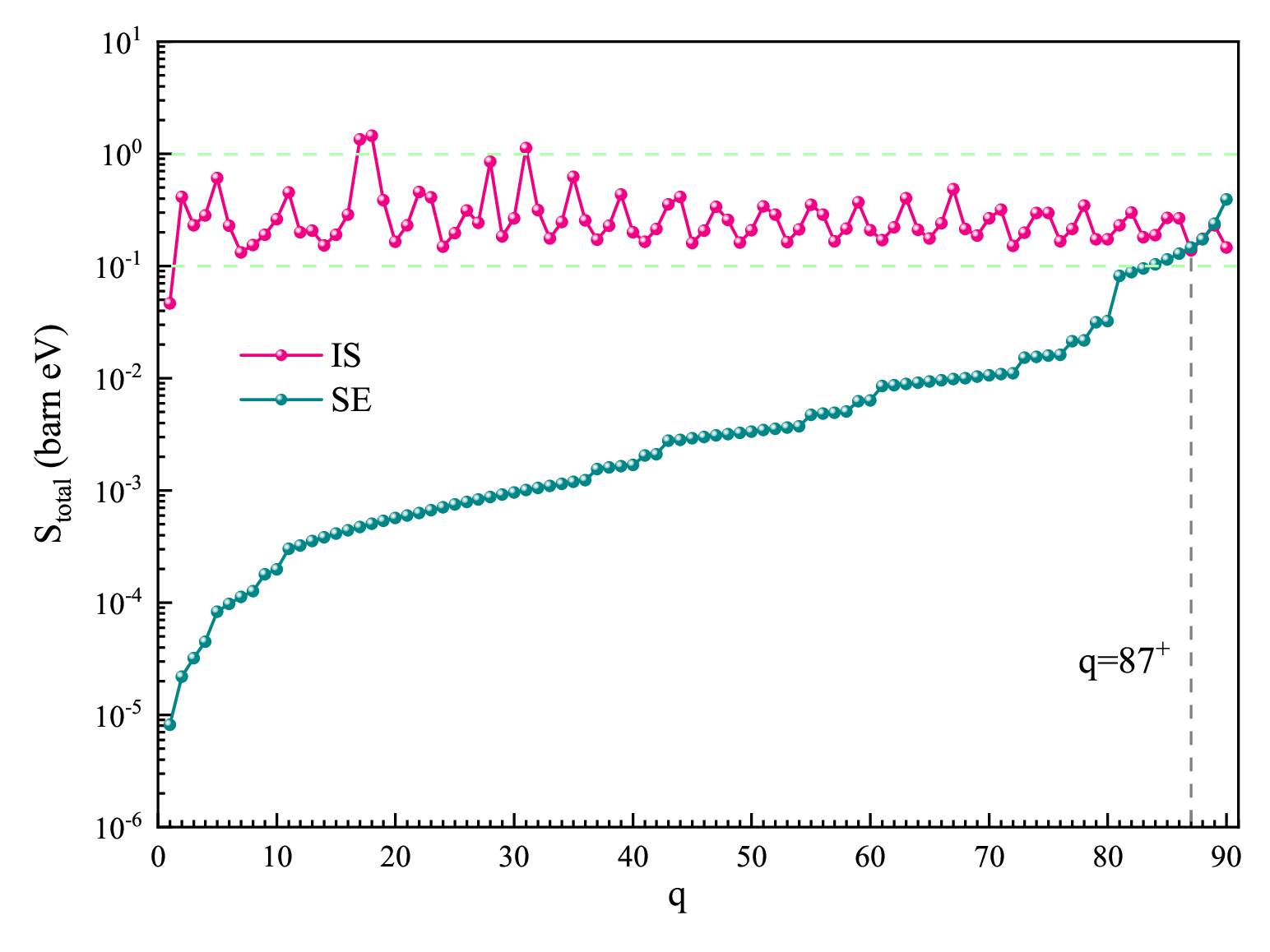}
	\caption{Total resonance strength $S_{\text{total}}$ for IS and SE states as a function of charge state $q$.}
	\label{fig 101010}
\end{figure} 

These results reveal $q$'s dual role: for low excitation energies (IS), it acts as a channel filter, with coupling enhancement compensating for reduced channel quantity; for high excitation energies (SE), it functions as a coupling amplifier, enabling unconstrained efficiency growth. Finally, global trends in the total resonance strength $S_{\text{total}}$ are summarized in Fig. \ref{fig 101010}. As observed, the IS $S_{\text{total}}$--reflecting the regulatory balance between threshold migration and compensatory enhancement--remains remarkably stable, fluctuating within the range of $10^{-1}$ to $10^0$ $\text{barn eV}$. In contrast, the SE $S_{\text{total}}$ grows exponentially with $q$, surging from $10^{-6}$ to $10^1$ $\text{barn eV}$. Notably, these trends intersect at $q=87^+$, beyond which the SE $S_{\text{total}}$ surpasses that of the IS. Hence, for charge states in the range $q = 87^+$ to $90^+$, indirect population of the IS via regulating the SE's NEEC process is more feasible than direct excitation of the IS, which is limited by the stability of channel screening and coupling compensation. This indirect strategy offers two key advantages: first, it circumvents challenges in direct excitation, such as the migration of IS channels to the hard-to-regulate $n>115$ region under high charge states; second, it lowers requirements on the energy precision for lasers or electron beams. Consequently, it provides a practical parameter window for controllable, high-intensity IS population and facilitates parameter optimization in nuclear optical clocks.

\section{Summary} 
\label{Sec.IV}
Given NEEC's potential for controlling $^{229}\text{Th}$ nuclear states, this work examines NEEC in $^{229}\text{Th}^{q+}$ ions with a focus on the effect of quantum numbers $n, l, j$ and charge states $q$. It is shown that quantum numbers exert distinct effects on NEEC key parameters: the resonance strength $S$ decreases with increasing $n$, larger $l$ suppresses both $\sigma$ and $S$, and $j$ weakly affects $\Gamma_{\text{NEEC}}$ but strongly modulates $\sigma$. For the isomeric state, valid channels show threshold migration, with dominant $n$ increasing linearly with $q$ and single-$n$-channel $S$ stabilizes via compensatory nucleus-electron coupling to keep $S_{\text{total}}$ constant. For the second excited state, $S_{\text{total}}$ increases monotonically with $q$ due to negligible screening. 
Furthermore, we find that when $q \geq 87^+$, the total resonance strength $S_{\text{total}}$ of the SE prevails. This not only offers a new experimental path to indirectly populate the IS by regulating the SE but also clarifies charge-state-driven nuclear-electronic interactions, providing insights for the precise manipulation of $^{229}\text{Th}$ nuclear states.

\end{document}